\documentstyle[onecolumn]{mn}

\newcounter{parentequation}
\def\beglet{
  \addtocounter{equation}{1}%
  \setcounter{parentequation}{\value{equation}}%
  \setcounter{equation}{0}%
  \def\theequation{\arabic{parentequation}\alph{equation}}%
  \ignorespaces
}
\def\endlet{
  \setcounter{equation}{\value{parentequation}}%
  \def\theequation{\arabic{equation}}%
}

\def\ltsima{$\; \buildrel < \over \sim \;$}
\def\gtsima{$\; \buildrel > \over \sim \;$}
\def\simlt{\lower.5ex\hbox{\ltsima}}
\def\simgt{\lower.5ex\hbox{\gtsima}}

\def\etal{{\it et al.}\rm}
\def\etals{{\it et al. }\rm}
\def\mk2{\mu {\rm K}^2}

\begin{document}

\title[Destriping Errors]
{Effects of Destriping Errors on Estimates of the CMB Power Spectrum}

 \author[G. Efstathiou]{G. Efstathiou\\
Institute of Astronomy, Madingley Road, Cambridge, CB3 OHA.}

\maketitle

\begin{abstract}
Destriping methods for constructing maps of the Cosmic Microwave
Background (CMB) anisotropies have been investigated extensively in
the literature. However, their error properties have been studied in
less detail. Here we present an analysis of the effects of destriping
errors on CMB power spectrum estimates for {\it Planck}-like scanning
strategies. Analytic formulae are derived for certain simple scanning
geometries that can be rescaled to account for different detector
noise.  Assuming {\it Planck}-like low-frequency noise, the noise
power spectrum is accurately white at high multipoles ($\ell \simgt
50$). Destriping errors, though dominant at lower multipoles, are
small in comparison to the cosmic variance. These results show that
simple destriping map-making methods should be perfectly adequate for the
analysis of {\it Planck} data and support the arguments given in
an earlier paper  in favour of applying a fast hybrid power spectrum
estimator to CMB data with realistic `$1/f$' noise.

\vskip 0.1 truein


\vskip 0.3 truein

\end{abstract}

\section{Introduction}

The problem of constructing a map of the CMB anisotropies from a set
of time-ordered data (TOD) has been studied by many authors.  The
methods can be broadly divided into two classes: `optimal' methods
that provide a least squares map making solution ({\it e.g.} Wright
1996; Wright \etals 1996; Tegmark 1997a,b; Borrill \etals 2001; Natoli
\etals 2001; Dor\'e \etals 2001) and approximate `destriping' methods
({\it e.g.}  Burigana \etals 1997; Delabrouille 1998; Maino \etals
1999; Revenu \etals 2000; Keih\"anen \etals 2003). A brute force
application of `optimal' methods requires the inversion of large
matrices and is computationally impractical for large TODs such as
those expected from WMAP and  {\it Planck}\footnote{See the {\it
Planck} web-site http://www.rssd.esa.int/index.php?project=PLANCK.}. As a result,
iterative algorithms have been developed ({\it e.g.} Wright \etals
1996; Natoli \etals 2001; Dor\'e \etals 2001) which do not require
matrix inversions. Nevertheless, for {\it Planck}-sized datasets,
these iterative algorithms are computationally expensive and require
the use of supercomputers.

Destriping algorithms are well suited to a {\it Planck}-type scanning
strategy in which the sky is scanned many times on rings.  The TOD can
then be averaged on rings and the effects of low frequency noise noise
approximated by a constant offset for each ring. The offsets can be
determined from the ring overlaps.  Destriping algorithms are
conceptually simple and computationally fast. Even for {\it
Planck}-sized TODs it is practical to apply destriping map making
methods on many thousands of simulations to test the effects of
various systematic errors (see {\it e.g.} Poutanen \etals 2004).

The motivation for this paper is twofold. Firstly, although a number
of authors have investigated destriping algorithms, almost all of this
work has been numerical. A notable exception is the paper by Stompor
and White (2004) who present an analysis of destriping errors for some
simple scanning strategies. One of the aims of this paper is to
develop on the work of Stompor and White and to derive an analytic
model of the effects of destriping errors on the CMB power spectrum
for {\it Planck}-like scanning strategies. This is useful
because it helps in developing an understanding of the map-making
process and how the errors depend on the parameters of the
experiment. This analysis also sheds light on the differences between
destriping and `optimal' map-making methods. In particular, whether the
extra computational cost and complexity of an optimal method actually
produces any significant improvement on simple destriping.

The second motivation for this paper follows from the need to compute
estimates of the CMB power spectrum, $C_\ell$, rapidly and acurately.
In an earlier paper (Efstathiou 2004, hereafter E04) a fast hybrid
estimator was developed that combines a quadratic maximum likelihood
estimator at low multipoles with a set of `pseudo-$C_\ell$' estimates
at high multipoles with different pixel weightings. In E04 this method
was tested against numerical simulations that used a realistic
scanning strategy for {\it Planck}, but assumed uncorrelated white
noise. In this approximation, the hybrid estimator was shown to be
very close to optimal and, importantly, can provide an accurate estimate
of the covariance matrix $\langle \Delta C_\ell \Delta C_{\ell^\prime}
\rangle$. However, in a realistic experiment, striping errors will 
introduce correlations in the noise. The question then arises as to
whether the hybrid estimator is applicable, for example, is there a
natural angular scale below which the noise can be assumed  white, and
if so, what fixes this scale?

Our goal, therefore, is to investigate the effects of destriping
errors on the CMB power spectrum for realistic scanning strategies and
noise models.  It should be emphasised that we do not attempt to
develop a new map making technique, nor to investigate `real world'
complexities such as asymmetric beams, positional errors or
non-stationary noise, though we will comment on how some of these
aspects may effect our results.

\section{Overview of Map Making with Destriping}

We denote the noise contribution to the TOD by $n(t)$, which is
considered to be a vector specified  at integer values of the
sampling frequency $t_{\rm samp}$, {\it i.e.} $n_i = n(it_{\rm
samp})$, $i=1, \dots N$.  The discrete Fourier transform of the noise
TOD is denoted $\hat n( \nu)$ and the power spectrum $\langle \vert
\hat n(\nu) \vert^2 \rangle$ is assumed to be of the form
\begin{equation}
P_n(\nu) = {\sigma^2_n \over T \nu_{\rm max}} \left ( 1 + {\nu_{\rm knee} \over \nu} \right )
, \qquad {1 \over T} \le \nu \le {1 \over 2 t_{\rm samp}},   \label{DS1} 
\end{equation}
where $\nu_{\rm knee}$ is the `knee frequency'and  $T$ is the total 
length of the time-stream.
The variance of the noise contribution given the power spectrum (\ref{DS1}) is 
\begin{equation}
\langle \vert n(t) \vert^2 \rangle = \sigma^2_n \left ( 1 + \nu_{\rm knee} {\rm \ln}(0.5T/t_{\rm samp})
\right ), \label{DS2}
\end{equation}
and thus  $\sigma^2_n$ fixes the overall amplitude of the noise and is equal
to the variance of white noise in the limit $\nu_{\rm knee} \rightarrow 0$.

The actual TOD is the sum of the true sky signal $s_p$ and the noise TOD
\begin{equation}
x_i = P_{ip}s_{p} + n_i,  \label{DS3}
\end{equation}
where $P_{ip}$ is a pointing matrix mapping a pixel $p$ on the sky to the pixel
$i$ in the TOD.
For a {\it Planck}-type scanning strategy, the same circle on the sky is mapped
$N_{\rm repoint}=60$ times before the satellite is repointed. We will denote the spin period
by $t_{\rm spin}$ and the repointing time interval by $t_{\rm repoint}$ ($\equiv N_{repoint}
t_{\rm spin}$). As explained in the introduction, the main aim of this paper is
not to develop an `optimal' map-making technique, but rather to gain
an  intuitive understanding
of the error properties of simple destriping map making methods. We therefore assume perfect
pointing and simply average the TOD on each scanning ring:
\begin{equation}
\bar x_i^k = \sum_{j=0}^{N_{\rm repoint}-1} x_{i + jN_{\rm ringpix} + (k-1)N_{\rm repoint}
N_{\rm ringpix}}, 
\qquad N_{\rm ringpix} \equiv t_{\rm spin}/t_{\rm samp}. \label{DS4}
\end{equation} 
where $N_{\rm ringpix}$ is the number of pixels withing a single ring pixel
and $\bar x_i^k$ is the mean vaue of the TOD in ring pixel $i$ of ring $k$.
This simple averaging is the maximum likelihood solution for map making on
a ring if the pointing is assumed to be perfect. It is possible  to improve on 
(\ref{DS4}) to account for imperfect pointing  by solving the maximum
likelihood equations to reconstruct a map from all scanning rings within a
single pointing period (see, for example, van Leeuwen \etals 2002).

\begin{table*}
\bigskip

\centerline{\bf \ \ \  Table 1:  Notation and parameters}

\begin{center}
\begin{tabular}{llr}   \hline \hline
{\rm symbol} & {\rm description} & {\rm value} \cr \hline \smallskip
$N_{\rm ring}$ & {\rm number}\ {\rm of} \ {\rm rings} & $2160$ \cr
$N_{\rm ringpix}$ & {\rm number}\ {\rm of}  \ {\rm pixels} \ {\rm per} \ {\rm ring}& $2160$ \cr 
$N_{\rm repoint}$ & {\rm number}\ {\rm of} \ {\rm rings} \ {\rm per} \ {\rm repointing} & $60$ \cr
$T$  & {\rm length }\ {\rm of} \ {\rm TOD}  & $7.78\;{\rm Ms}$  \cr
$t_{\rm repoint}$  & {\rm pointing }\ {\rm period}  & $3600\;{\rm s}$  \cr  
$t_{\rm spin}$  & {\rm spin }\ {\rm period}  & $60\;{\rm s}$  \cr 
$t_{\rm samp}$  & {\rm sampling}\ {\rm period}  & $27.78\;{\rm ms}$  \cr 
$\nu_{\rm spin}$  & {\rm spin }\ {\rm frequency}   & $16.67\;{\rm mHz}$  \cr 
$\nu_{\rm knee}$  & {\rm knee }\ {\rm frequency}   & $30.00\;{\rm mHz}$  \cr 
$\nu_{\rm max}$  & {\rm maximum }\ {\rm frequency}   & $18.00\;{\rm Hz}$  \cr 
$\sigma_N$  & {\rm noise}\ {\rm amplitude}   & $1.36\;{\rm mK}$  \cr 
$\Delta \theta_{\rm ring}$  & {\rm ring }\ {\rm width}   & $10\;{\rm arcmin}$  \cr 
$\Delta \theta_{\rm ringpix}$  & {\rm size} \ {\rm of} {\rm ring }\ {\rm pixel}   & $10\;{\rm arcmin}$  \cr 
$\Delta \theta_{\rm map}$  & {\rm size} \ {\rm of} {\rm map }\ {\rm pixel}   & $15\;{\rm arcmin}$  \cr 
$\theta_{\rm beam}$  & {\rm beam} \ {\rm width} {\rm FWHM}   & $30\;{\rm arcmin}$  \cr 
$N_{\rm map}$ & {\rm number} \ {\rm of} \ {\rm map} \ {\rm pixels}  & $659676$  \cr
$x_i$ & {\rm signal}+{\rm noise} \ {\rm in} \ {\rm TOD} \ {\rm  pixel}  $i$ & \cr
$\bar x_i^k$ & {\rm mean} \ {\rm signal}+{\rm noise} \ {\rm in} \ {\rm  pixel}  $i$ { \rm of} \ {\rm ring}  $k$ & \cr
$\epsilon_k$ & {\rm offset} \ {\rm of} \ {\rm ring}  $k$ & \cr
$m_p$ & {\rm map} \ {\rm pixel}   $p$ & \cr
\hline
\end{tabular}
\end{center}  

\end{table*}
\medskip

For the simulations presented here,  we adopt
the parameters listed in Table 1 unless  stated otherwise.
Table 1 also serves as a summary of
the notation used in this paper. The knee frequency in Table 1 has
been chosen to be representative of the $70$ GHz channel of the {\it
Planck} Low Frequency Instrument (LFI) (see Tuovinen 2003) . This is
an interesting case because the knee frequency is about twice the spin
frequency. The knee frequencies for the {\it Planck} High Frequency
Instrument (HFI) should be smaller than the spin frequency. This case
is less interesting because if $\nu_{\rm knee} \ll \nu_{\rm spin}$, it
is a very good approximation to model the low frequency noise as a
constant offset in each ring. The noise level $\sigma_n$ for the
simulations has been chosen so that the CMB power spectrum is noise
dominated at multipoles $\ell \simgt 300$ (rather than to match the
noise for any of the Planck detectors).  The input CMB power spectrum, 
$C_\ell$,  is that of the concordance $\Lambda$CDM model favoured by
WMAP (Spergel \etals 2003). As in E04, unless stated otherwise beam
functions will not be written explicitly in equations, thus $C_\ell$
will ususally mean $C_\ell b_\ell^2$, where $b_\ell$ is the spherical
transform of a symmetric Gaussian beam. The remaining parameters, such
as the number of rings, number of ring pixels, map pixel size {\it
etc} were chosen so that large numbers of simulations could be run
quickly.

The basic assumption behind destriping techniques is that  low
frequency drifts in the TOD can be accounted for by adding a 
constant offset $\epsilon_k$  to each ring. The ring offsets can
be determined by minimising
\begin{equation}
S = \sum_{kl j\subset i}  (\epsilon_k + \bar x_i^k - \epsilon_l - \bar
x_j^l)^2 + \lambda \left ( \sum_l \epsilon_l \right)^2, \label{DS5}
\end{equation}
where the notation $j \subset i$ indicates that the ring pixels $j$ and $i$
overlap the same map pixel $m_p$. The second term in equation (\ref{DS5}) is included
to enforce the condition $\sum_k \epsilon_k = 0$. The offsets are therefore
given by the solution of the linear equations
\begin{equation}
\sum_{l j\subset i}  (\epsilon_k + \bar x_i^k - \epsilon_l - \bar
x_j^l) + \lambda \left ( \sum_l \epsilon_l \right) = 0. \label{DS6}
\end{equation}
There has been some discussion in the literature concerning the
weighting of the term in the first summation in equation (\ref{DS5}).
Equation (\ref{DS5}) assigns equal weight to each overlapping pixel,
as in Maino \etal (1999). Delabrouille (1998) assigns a weight $w=
1/(n_p-1)$, where $n_p$ is the total hit count in map pixel $p$, while
Keih\"anen \etals (2003) assign a weight $w = 1/n_p$. For {\it
Planck}-like scanning strategies the differences between destriping
using these weight functions are much smaller than the striping errors
themselves (see Figure 3 of Keih\"anen \etals 2003) and so the weight
function will be set to unity throughout this paper.

Equations (\ref{DS6}) can be solved by a matrix inversion and the solution
is independent of the regularising parameter $\lambda$, provided it is chosen
to be large enough. Alternatively, these equations can be solved by iteration by
setting
\begin{equation}
\epsilon_k  \approx  {1 \over N_{ok}} \sum_{l j\subset i}  (\bar x_j^l - \bar x_i^k),  \label{DS7}
\end{equation}
adding the derived offsets to $\bar x_i^k$ and re-evaluating equation
(\ref{DS7}) until the offsets (\ref{DS7}) converge to zero. In (\ref{DS7}),
$N_{ok}$ is the number of pixels on all rings that overlap
pixels on ring $k$. This algorithm is
identical to the plate matching procedures applied to create galaxy
catalogues from photographic data (Groth and Peebles 1986; Maddox,
Efstathiou and Sutherland 1996). If the number of overlaps per ring is
large, then an accurate estimate of the variance of the offsets can be derived
from the first iteration of equation (\ref{DS7}),
\begin{equation}
\langle \epsilon_k^2 \rangle   \approx  {(N_{oll^\prime\subset k} + N_{ok}) \over N_{ok}^2} {\sigma^2_n \over N_{repoint}} \approx 
{N_{oll^\prime\subset k}  \over N_{ok}^2}{\sigma^2_n \over N_{\rm repoint}},  \label{DS8}
\end{equation}
where $N_{oll^\prime \subset k}$ is the number of identical ring
pixels summed over all pairs of rings $l$, $l^\prime$ that overlap
with ring $k$. In equation (\ref{DS7}) we have assumed that the noise
in each averaged ring pixel is white, with variance $\sigma^2_n/N_{\rm
repoint}$, which is accurate on a single ring even in the presence of
$1/f$ noise unless $\nu_{\rm knee} \gg \nu_{\rm spin}$. The second
term on the right hand side of equation (\ref{DS8}) applies for a {\it
Planck}-type scanning strategy for which $N_{oll^\prime \subset k} \gg
N_{ok}$.

We consider simple {\it Planck}-type scanning strategies with the spin
axis either aligned with the ecliptic plane, or with a slow precession
of $5^\circ \sin(2 \phi_e)$ about the ecliptic plane, where $\phi_e$ is the ecliptic
longitude.  The sky is scanned with a single detector at a`
bore-sight' angle of $\theta_b$ with respect to the spin axis. After a
complete uniform sweep of the ecliptic plane, the time-stream maps to
a set of $N_{\rm rings}$ each of angular width $\Delta \theta_{\rm
ring}$.

\begin{figure*}
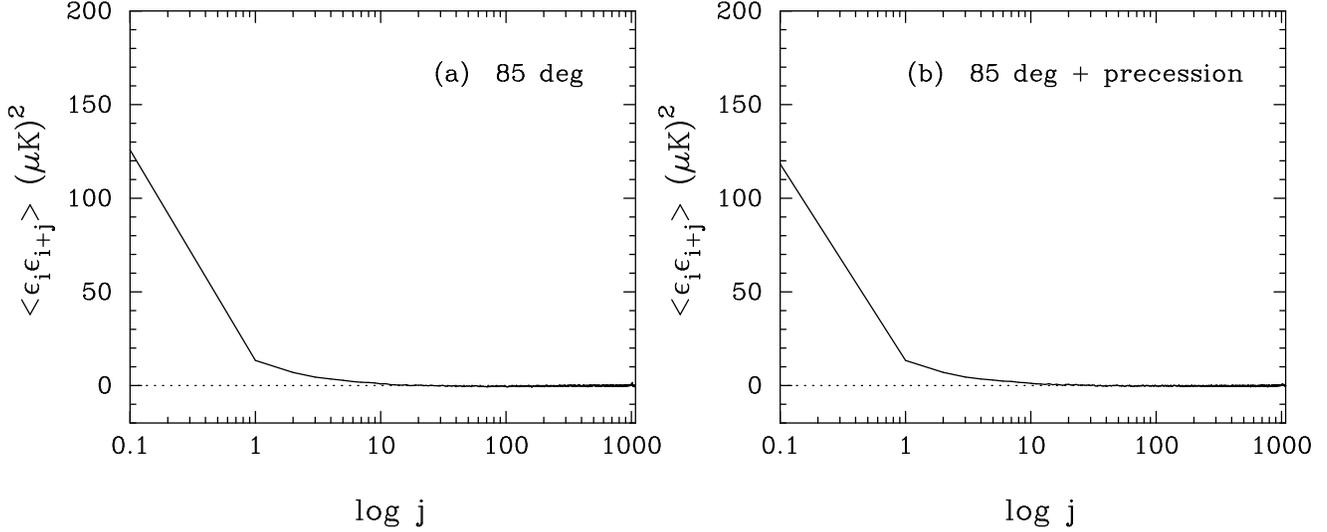


\vskip 3.5 truein

\includegraphics{pgc12a.ps}
\includegraphics{pgc12b.ps}

\caption
{The correlation function of the ring offsets for two {\it Planck}-type
scanning strategies: (a) for a bore-sight angle $\theta_b = 85^\circ$ and 
spin axis aligned with the ecliptic plane; (b) for $\theta_b = 85^\circ$ and with
a slow sinusoidal precession of $\theta = 5^\circ$ above and below the ecliptic
as discussed in the text. The indices $i$ and $j$ refer to the
ring number.}

\label{figure1}

\end{figure*}

Figure 1 shows the correlation functions of the ring offsets averaged over
$250$ simulations for each of the two scanning strategies discussed in the 
previous paragraphs. There figures indicate the following:

\smallskip

\noindent
(a) The dispersion in the ring offsets is $\langle \epsilon_i^2
\rangle^{1/2} \approx 11 \;\mu {\rm K}$\footnote{For {\it
Planck} the dispersion will be smaller because the detector noise is
smaller than assumed here} in excellent agreement with equation
(\ref{DS8}). This is much smaller than the white noise
level of $176 \; \mu {\rm K}$ on a single ring because the number of
overlaps for a {\it Planck}-type scanning strategy is large. As
emphasized by Stompor and White (2004) the pixel noise of the
resulting maps will be predominantly white and uncorrelated.

\smallskip

\noindent
(b) It is interesting to compare the ring variance with the expected signal variance
for the case of no precession:
\begin{equation}
\langle \epsilon_k^2 \rangle = {1 \over 4 \pi} \sum_\ell (2 \ell + 1)
C_\ell \left [ P_\ell ( \rm cos \; \theta_b ) \right ]^2 = (14.2 \; \mu
{\rm K})^2.  \label{DS9}
\end{equation}
Thus, for the parameters adopted in this paper, the offset variance arising 
from ring pixel noise is comparable to the true signal variance.

\smallskip

\noindent
(c) The offset correlation functions in Figure (\ref{figure1}) drop
rapidly to zero. To high accuracy, we can model the striping errors as
a set of random offsets with dispersion $\sigma$ and effective
ring-width $\Delta \alpha$ as illustrated in Figure 2. As shown in the
next Section, The contributions of these errors to the spherical
harmonics and power spectrum of the destriped maps can, in this
approximation, be calculated exactly for the case of a perfect ring
torus.

\smallskip

\noindent
(d) There is almost no perceptible difference in the correlation
functions for the two scanning strategies. We would therefore expect
(and this is verified in Section 4) that the effects of destriping
errors on the power spectra should be very similar for these two scanning
strategies.

\smallskip

\noindent
(e) The striping errors arising from the dispersion in the ring
offsets are `irreducible' errors. By this, we mean that these errors
are fixed by the instrumental white noise on the rings and the
crossing points (interconnectedness) of the TOD. They cannot be
reduced by applying more time-consuming `optimal' map-making methods
(see Section 5).
However, since the knee frequency $\nu_{\rm knee}$ exceeds both the
repointing frequency $1/t_{\rm repoint}$ and the spin frequency, the
averaged ring data $\bar x_i^k$ will contain low amplitude gradients
associated with `$1/f$' noise.  These gradients can, in principle, be
removed by modifying the destriping code to determine from the crossing points 
a few low order coefficients in, say, a Legendre poynomial or Fourier expansion,
(Delabrouille 1998; Keih\"anen \etals 
2003) However, as we will show in Section
3, the effects of these gradients are smaller than the effects of the 
offset errors.

\begin{figure*}

\vskip 3.8 truein

\includegraphics{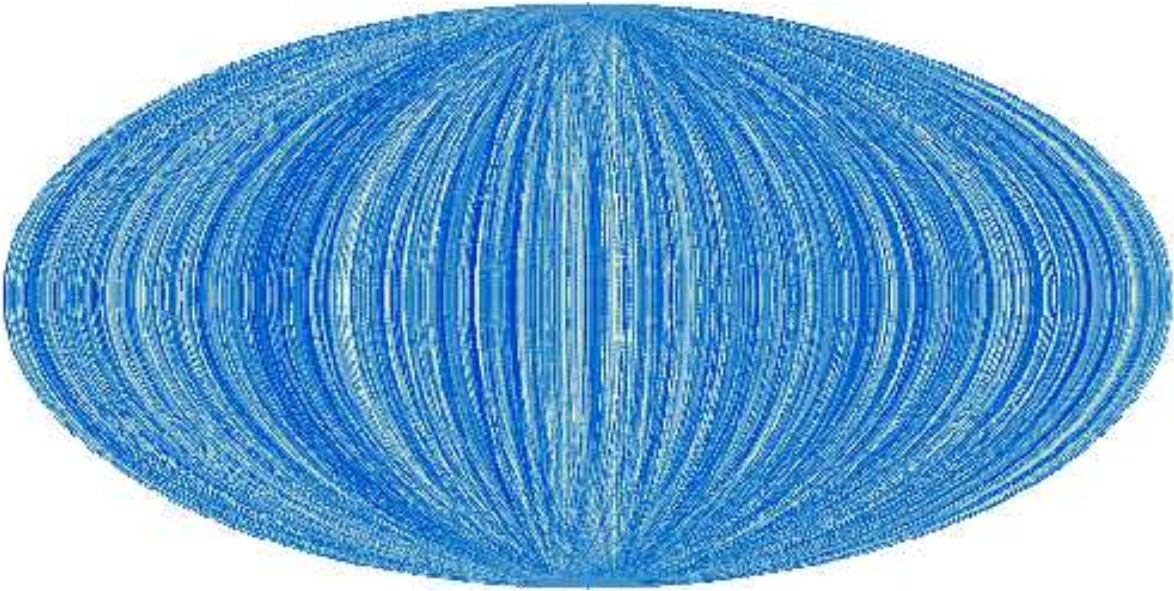}

\caption
{A map of random striping errors for a ring torus with $\theta_b =
85^\circ$. Note that there is some `ringing' with the map
pixelisation, but as will be shown in Section 3 this ringing has a
negligible effect on the power spectrum of the ring offsets.}

\label{figure2}

\end{figure*}

\section{Analytic Models of Destriping Errors}

In this Section we consider a scanning strategy which leads to a perfect
ring torus, {\it i.e.} the spin axis is aligned with  the ecliptic plane
as the sky is scanned by a single detector with a bore-sight angle of
$\theta_b$. For such a scanning strategy, the distribution of 
hit-counts on the sky will follow a distribution with 
ecliptic latitude of
\begin{equation}
  dH(\theta) =  \left\{ \begin{array}{cc} 
                 N  (1 - \cos^2 \theta/\sin^2\theta_b )^{-1/2}\sin \theta \; d\theta,
                 &  \cos \theta  \le \sin \theta_b , \\
                 0, &   \cos \theta  > \sin \theta_b . \end{array} \right.
 \label{RT1}
\end{equation}
The hit-count distribution is therefore highest at $\cos \theta = \sin \theta_b$
and lowest at the ecliptic $\theta = \pi/2$.

As described in the previous Section, the offsets $\epsilon_k$ of each
ring can be modelled as a set of independent Gaussian random variates with
dispersion $\sigma$.  The map constructed from these ring offsets
(which we will refer to as the error map) will contain most of the
information on pixel correlations introduced by the map-making
process. Our goal in this Section is to compute the effects
of these errors on the CMB power spectrum.

The spherical harmonic transform of the error map can be written as
\begin{equation}
 a^e_{\ell m} = \sum_{ik} w_{ik}\Delta T_{ik}
\Omega_{ik} Y_{\ell m}( \theta_{ik}, \phi_{ik}),     \label{RT2}
\end{equation}
where the index $k$ denotes the ring number, $i$ denotes the
pixel number within the ring and $\Omega_{ik}$ is the solid angle
of the ring pixel. The weight factors $w_{ik}$ account for the 
averaging of the ring pixels in constructing the map and thus are
proportional to the inverse of the hit count distribution of equation
(\ref{RT1}).

To evaluate equation (\ref{RT2}), reorient each ring to a new coordinate
system $(\theta^\prime, \phi^\prime)$ in which the spin axis is aligned
with the new $z^\prime$ axis. The spherical harmonic transform for each ring
is then,
\beglet
\begin{equation}
 a^{\prime ek}_{\ell m}  = {1 \over 2} \sigma_k A_{\ell}^{m} P_{\ell}^{m}(\cos \theta_b)
\sin \theta_b \Delta \alpha \int_0^{2 \pi} \vert \sin \phi^\prime \vert 
{\rm e}^{i m \phi^\prime} d\phi^\prime,    \label{RT3a}
\end{equation}
where $\sigma_k$ is the constant ring-offset $\Delta T_{ik}=\sigma_k$ and
the $A_{\ell}^{m}$ are the normalising factors of the spherical harmonics
\begin{equation}
A_{\ell}^{m} = \left ( {2 \ell + 1 \over 4 \pi} {( \ell - m)! \over (\ell + m)!} 
\right )^{1/2}. \label{RT3b}
\end{equation}
\endlet
Notice that  the hit count distribution  has been normalised
so that the sum of the weight factors
\begin{equation}
\sum_{ik} w_{ik} \Omega_{ik} = 4 \pi \sin \theta_b \nonumber
\end{equation}
for a complete ring torus (which completely covers the sky twice for
the case $\theta_b = \pi/2$).
Performing the integral over $\phi^\prime$ in equation (\ref{RT3a}),
\begin{equation}
 a^{\prime ek}_{\ell m}  =  \sigma_k  \sin \theta_b \Delta \alpha
A_\ell^m P_\ell^m
(\cos \theta_b) {1 \over (1 - m^2)} ( 1 + (-1)^m)
=  \sigma_k  \sin \theta_b \Delta \alpha K_{\ell}^m(\cos \theta_b), \label{RT4}
\end{equation}
where $K_{\ell}^{m}$ is zero for all odd values of $m$.
Transforming back into the original coordinate system
\begin{equation}
a^{e}_{\ell m} = \sum_{k} \sum_{m^\prime}   a^{\prime ek}_{\ell m\prime}D^\ell_{m^\prime}
(\alpha_k, \beta_k, \gamma_k), \label{RT5} 
\end{equation}
where the $D^{\ell}_{mn}$ are the Wigner D-matrices (see {\it e.g.} Brink
and Satchler 1993; Varshalovich, Moskalev and Khersonskii 1988) and $\alpha$,
$\beta$ and $\gamma$ are the Euler angles relating the two coordinate
systems. In our case, $\alpha_k=0$, $\beta_k = \pi/2$, hence in terms
of the real reduced rotation matrices (\ref{RT5}) is
\begin{equation}
a^{e}_{\ell m} = \sum_{k} \sum_{m^\prime}  
\sigma_k \Delta \alpha \sin \theta_b K_\ell^{m^\prime} d^{\ell}_{m^\prime m} (\pi/2) {\rm e}^{-im\gamma_k}. \label{RT6} 
\end{equation}

\begin{figure*}
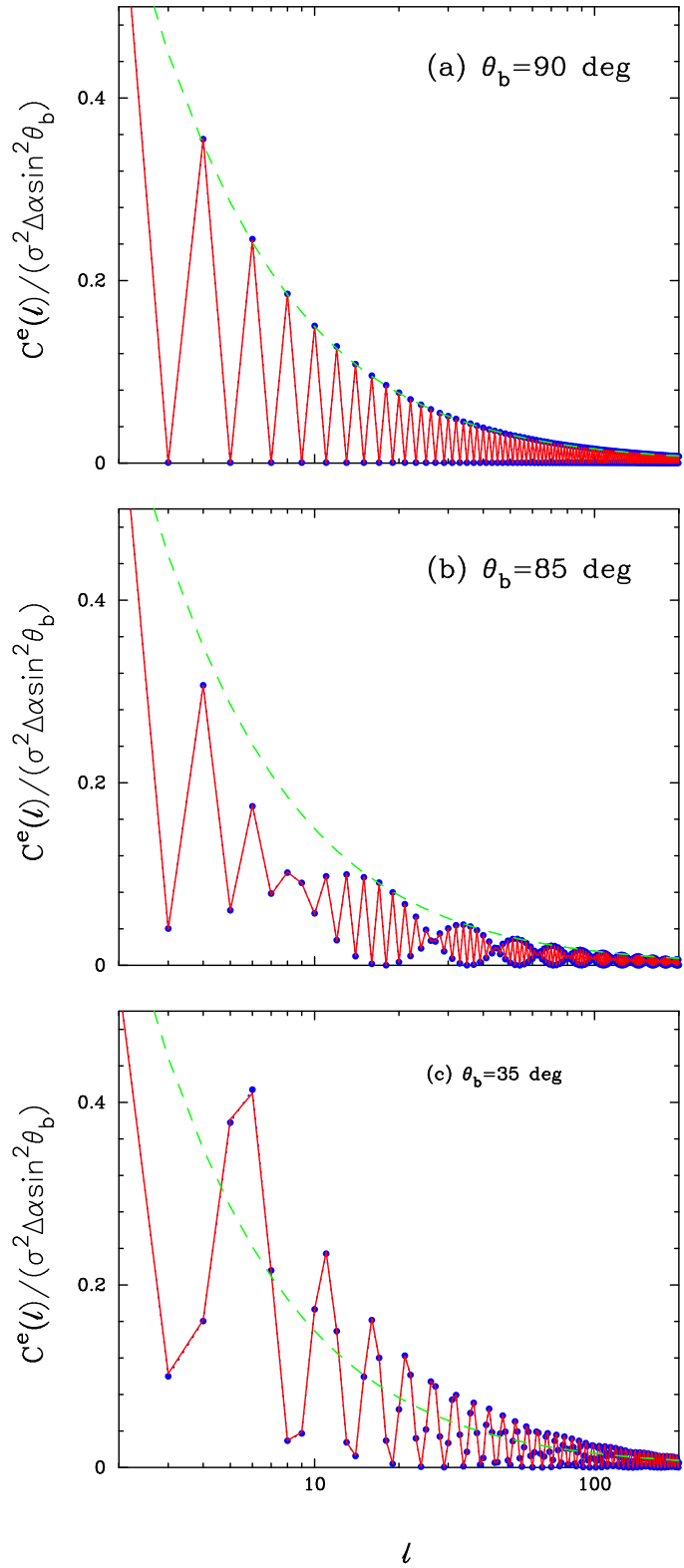


\vskip 8.5 truein

\includegraphics{pgclerror_1.ps}
\includegraphics{pgclerror_2.ps}
\includegraphics{pgclerror_3.ps}

\caption
{The power spectrum of destriping errors for ring tori with various 
values of $\theta_b$. In each figure the filled (blue) points show the results
from numerical simulations and the solid (red) lines show the analytic 
expression of equation (\ref{RT9}). The dot-dashed (green) line
shows the simple analytic approximation of equation (\ref{RT12}).}

\label{figure3}

\end{figure*}

Defining the power spectrum of the error map as
\begin{equation}
\tilde {C}^e_{\ell} = {1 \over (2 \ell + 1)} \sum_{m} \langle a^e_{\ell m}
a^{e*}_{\ell m}, \rangle \label{RT7}
\end{equation}
(where the tilde on $\tilde C^e$ signifies that the power spectrum is computed
from the $a^e_{lm}$ coefficients computed on the incomplete sky if $\theta_b \ne \pi/2$)
and using the relation
\begin{equation}
\sum_{m} d^\ell_{m^\prime m} (\pi/2) d^\ell_{m^{\prime \prime} m} (\pi/2) = \delta_
{m^{\prime} m^{\prime \prime}}, \label{RT8}
\end{equation}
we find
\begin{equation}
\tilde {C}^e_{\ell} = {2 \pi \over (2 \ell + 1)} \sigma^2 \Delta \alpha
\sin^2 \theta_b \sum_{m} \vert K_\ell^m (\cos \theta_b)\vert^2. \label{RT9}
\end{equation}

For the special case $\theta_b = \pi/2$, the factor $\vert K_\ell^m \vert$
is proportional to $P_\ell^m(0)^2$ and so vanishes for odd values of 
$(\ell -m)$. Since $K_{\ell}^m$ is zero for odd values of $m$, it follows
that $\tilde C^e_{\ell} = 0$ for odd values of $\ell$. An alternative 
derivation of the $\tilde {C^e_\ell}$ for the special case $\theta_b = \pi/2$,
(in which the spherical harmonic transform for each ring is evaluated in a
coordinate system with the $z$-axis perpendicular to the spin axis)
gives
\beglet
\begin{equation}
\tilde {C}^e_{\ell} = {1 \over 2} \sigma^2\Delta \alpha I_\ell, \label{RT10a}
\end{equation}
where
\begin{eqnarray}
  I_\ell = \left\{ \begin{array}{ccc}
 -\int_0^\pi \theta \cos\theta P_\ell (\cos \theta) d\theta 
                 &  \ell \;\; {\rm even}, &  \ell \ne 0 \\
                 0, &   \ell \;\; {\rm odd}. &  \end{array} \right.
 \label{RT10b}
\end{eqnarray}         
\endlet
For large values of $\ell$ the integral in equation (\ref{RT10b}) can be
approximated by 
\begin{equation}
 -\int_0^\pi \theta \cos\theta P_\ell (\cos \theta) d\theta \approx \pi \int_0^\infty
J_0 ((\ell + 1/2) \theta) = {\pi \over (\ell + 1/2)}, \label{RT11}
\end{equation}         
hence
\begin{equation}
\tilde {C}^e_{\ell} \approx {\pi \over (2\ell + 1)} \sigma^2\Delta \alpha, 
\qquad \ell \;\; {\rm even}.
\label{RT12}
\end{equation}
Equation (\ref{RT12}) is, in fact, an extremely good approximation (to within 2\%)
to the exact answers of equations (\ref{RT9}) and (\ref{RT10a}) for values of 
$\ell$ as small as $\ell=4$.

Some examples of the error power spectra for three values of
$\theta_b$ are shown in Figure \ref{figure3}. For each value of
$\theta_b$, $10^4$ simulations were generated each with $720$ rings
assigned random offsets. The rings were then mapped on to the igloo
pixelization scheme described in E04 with $0.25^\circ \times
0.25^\circ$ pixels simply by assigning the nearest map pixel to each
ring pixel and averaging over all of the ring pixels assigned to any
particular map pixel. The power spectra were computed from the igloo
maps using using fast spherical transforms.  As can be seen from
Figure \ref{figure2}, the finite sizes of the ring and map pixels
introduce some structure in the final error maps.  Nevertheless, the mean
power spectra for the error maps (shown by the points in Figure
\ref{figure3}) agree perfectly with the analytic results of equation
(\ref{RT9}) which were derived in the continuum limit. Evidently, the
effects of finite pixelisation and ring widths are negligible and
hence the analytic model developed in this Section gives an extremely
accurate representation of the destriping errors.

\begin{figure*}
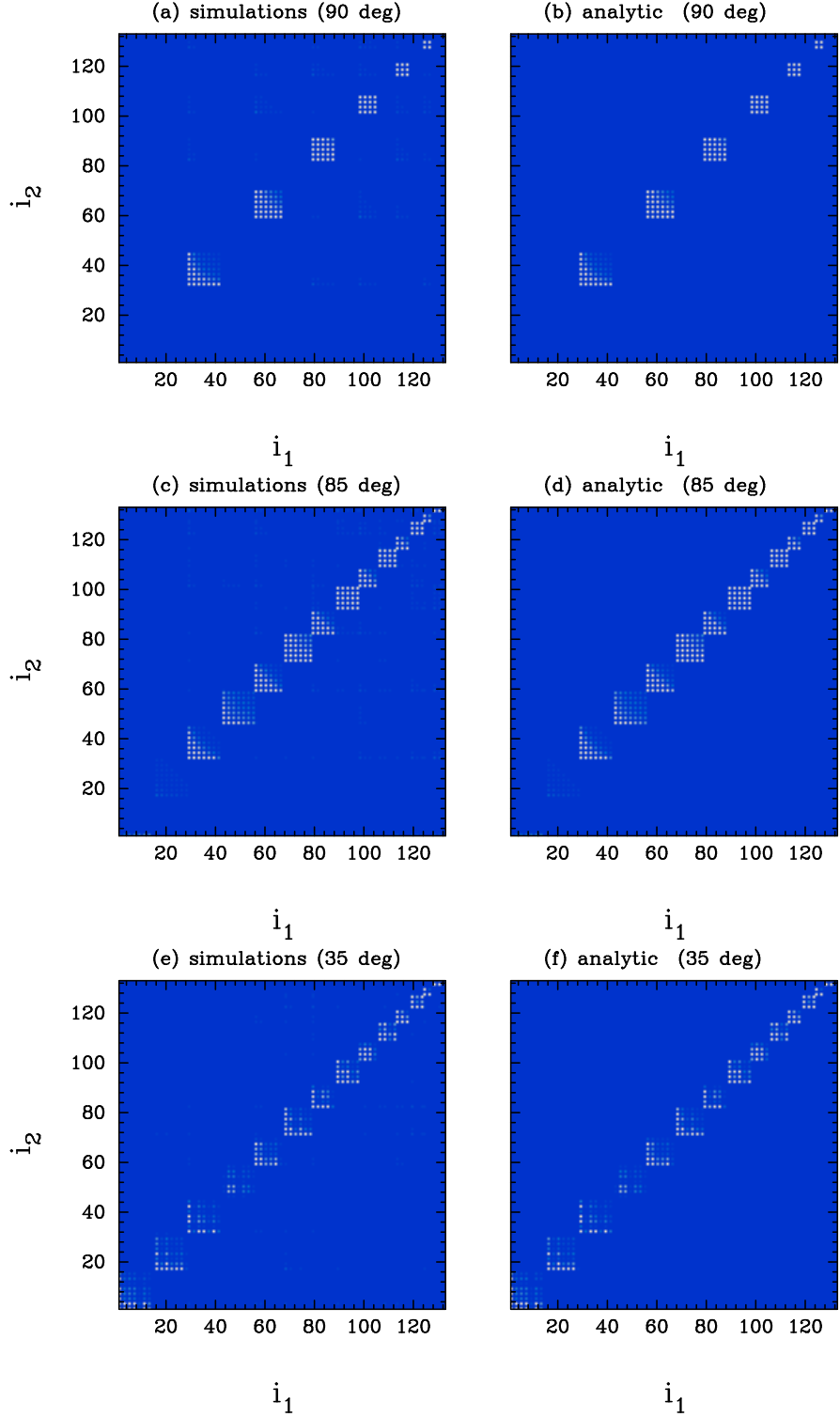


\vskip 8.1 truein

\includegraphics{cl_cov_90.ps}
\includegraphics{cl_cov_85.ps}
\includegraphics{cl_cov_35.ps}

\caption
{The covariance matrices for $\langle a^e_{\ell m} a^{e*}_{\ell^\prime
m^\prime} \rangle$ for the simulations shown in Figure (\ref{figure3})
compared to the analytic expression of equation (\ref{RT13}). The indices
$i_1$ and $i_2$ are ordered as $(m, \ell)$, 
({\it i.e.} $(0, 0), \dots (0, \ell_{\rm max})$, 
$(1, 0), \dots (1, \ell_{\rm max})$, {\it etc})
with $m$ and $\ell$ running from $0$ to $10$.}

\label{figure4}

\end{figure*}

If the primordial  fluctuations are Gaussian, the spherical harmonic 
coefficients will satisfy 
\begin{equation}
\langle a^e_{\ell m} a^{e*}_{\ell^\prime m^\prime} \rangle
= C_\ell \delta_{\ell \ell^\prime} \delta_{m m^\prime}. \label{RT14}
\end{equation}
The striping erros will, however, introduce correlations in the $a_{\ell m}$.
From equation (\ref{RT6}) it is straightforward to show that the correlations
introduced by striping are given by 
\begin{equation}
\langle a^e_{\ell m} a^{e*}_{\ell^\prime m^\prime} \rangle
= 2 \pi \sigma^2 \sin^2 \theta_b \Delta \alpha \; \delta_{m m^\prime}
\sum_{m_1 m_2} K^{m_1}_\ell K^{m_2}_{\ell^\prime} d^{\ell}_{m_1 m}(\pi/2) 
d^{\ell^\prime}_{m_1 m^\prime}(\pi/2). \label{RT13}
\end{equation}
For the special case $\theta_b = \pi/2$, these correlations can be
written as
\beglet
\begin{equation}
\langle a^e_{\ell m} a^{e*}_{\ell^\prime m^\prime} \rangle
= 2 \pi \sigma^2  \Delta \alpha \; \delta_{m m^\prime}
E^m_\ell E^{m^\prime}_{\ell^\prime} (1 + (-1)^m), \label{RT14a}
\end{equation}
where
\begin{equation}
E^m_\ell = \int_{-1}^{1} A_\ell^mP^m_{\ell}(\mu) \;d\mu. \label{RT14b}
\end{equation}
\endlet 
If the vector $a^e_{\ell m}$ is ordered as $(m, \ell)$, 
({\it i.e.} $(0, \ell_{\rm min}), \dots (0, \ell_{\rm max})$, 
$(1, \ell_{\rm min}), \dots (1, \ell_{\rm max})$, {\it etc})
 the covariance matrix $\langle a^e_{\ell m}
a^{e*}_{\ell^\prime m^\prime} \rangle$ will have a block diagonal
structure. Figure \ref{figure4} shows examples of these covariance
matrices for the three scanning strategies discussed above. 
As we will show in the next Section, these off-diagonal
correlations will  be much smaller than the diagonal
components for the parameters of a realistic experiment.
However, since the departures from Gaussianity in most inflationary 
models are expected to be small (for a comprehensive review see Bartolo
\etals 2004), residual map-making errors could prove problematic in testing
for non-Gaussianity.
For the remainder of this paper, we will concentrate on errors on
estimates of  the CMB power spectrum, though it is important to 
recognise that even if map-making errors can be shown to have a
negligible effect on the power spectrum, they may be important
for other statistical tests.

\section{Simulations with Realistic Noise}

In this Section, we describe the results from a set of simulations
with realistic `$1/f$' noise (equation \ref{DS1}) with
the parameters given in Table 1. As explained in Section 2, 
the resolution and sizes of the maps
and ring sets were chosen so that large numbers of simulations could
be run quickly while demonstrating the salient features of the
map-making problem. As a further speed up, $1/f$-noise was generated
by using an FFT for frequencies below $0.133\;{\rm Hz}$.  Above this
frequency the noise was assumed to be white, which has the additional
advantage that the white noise can be added to ring pixels `on the
fly'  so that it is never necessary to store a complete TOD in
memory. A complete simulation, including noise generation, destriping
and power spectrum estimation takes approximately $70$ seconds on an
single $1.4\;$GHz Itanium 2 processor. Splitting the noise into a low
frequency `$1/f$' component and a high frequency white noise component
has the additional advantage that the effects of low and high
frequency noise on destriping can be investigated separately. 

\begin{figure*}
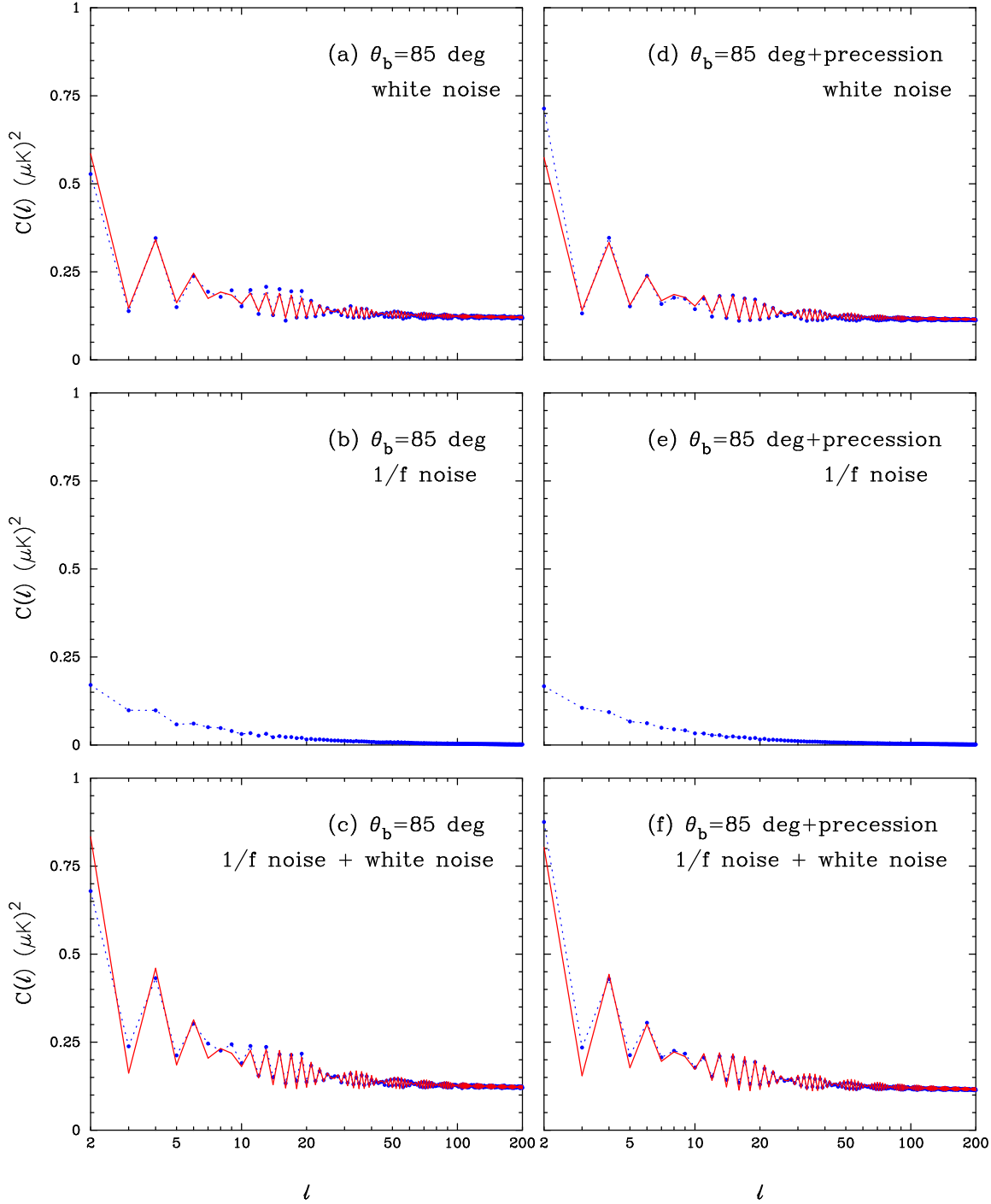


\vskip 7.5 truein

\includegraphics{pgsima.ps}
\includegraphics{pgsimb.ps}
\includegraphics{pgsimc.ps}

\includegraphics{pgsimd.ps}
\includegraphics{pgsime.ps}
\includegraphics{pgsimf.ps}

\caption
{Averages over $250$ simulations of noise for a single detector with
$\theta_b=85^\circ$ and no precession (Figures \ref{figure5}a, b, \&
c) and with a slow sinusoidal precession of $5^\circ$ amplitude as discussed in the
text (Figures \ref{figure5}d, e \& f). Filled circles in the upper
panels (a \& d) show residual errors after destriping assuming only
uncorrelated white noise on the rings. Filled circles in the middle
panels (b \& e) show residual errors after destriping assuming only
low frequency $1/f$ noise.  Filled circles in the  bottom panels (c \& f) 
show residual errors including white noise and low frequency $1/f$ noise.
The solid lines in the upper panels show the model for destriping errors
plotted in Figure 3b together with the white noise level appropriate to
the maps. The solid lines in the lower panels show the same destriping
model renormalised to fit the simulations.}

\label{figure5}

\end{figure*}

We have run two sets each of $250$ simulations for the two scanning
strategies adopted for Figure 1, namely, $\theta_b=85^\circ$ with no
precession and $\theta_b=85^\circ$ with a slow sinusoidal precession of $\pm
5^\circ$. In each case, we generated three ring-sets for each simulation:
one with white noise only, one with low frequency `$1/f$' noise only, and 
one using the sum of these two noise models. 
These ring-sets were passed through the destriping algorithm
to produce three maps per simulation. The averaged power spectra for
the two sets of simulations are plotted in Figure \ref{figure5}.

The upper panels in Figure \ref{figure5} show the power spectra for the case
of white noise only on the rings. This is the case that is closest to the
analytic model of the previous Section. As expected, the analytic model
of equation (\ref{RT9}) summed with the appropriate constant white noise
level,  provides an excellent match to the simulations
when proper allowance is made to calculate the effective ring width using
the correlation functions of Figure \ref{figure1}. This is true even for the 
slow precession scanning strategy shown in Figure \ref{figure5}d. The main
effect of a the slow precession is to fill in the coverage gaps at the 
eclipic poles, but the ring pattern is so close to a perfect ring torus that
the residual striping errors produce a nearly identical effect on the power
spectrum.

Figures \ref{figure5}b and \ref{figure5}e show the residual striping errors
when only low frequency `$1/f$' noise is included. These errors arise primarily
from residual gradients on the rings and so their amplitude  depends on the
knee frequency. Had we adopted a knee frequency much smaller than the spin
frequency, these errors would have had a much lower amplitude. Nevertheless,
even for the parameters adopted here, the amplitude of these errors
is considerable smaller than the errors caused by the dispersions in the ring
offsets. Notice that these errors also decay roughly as $1/\ell$, as for the
error power spectra for pure white noise. As mentioned earlier, Delabrouille
(1998) and  Keih\"anen \etals (2003) have explored fitting low order functions
for each ring during destriping, rather than a single offset $\epsilon_k$. The results
are mixed and in fact Keih\"anen \etals (2003) find that including more parameters
actually produces larger power spectrum errors than simply fitting constant offsets.
This is not suprising because the errors on the individual crossing points are
dominated by the white noise. If the white noise level is high, and the knee
frequency is low, there may not be enough crossing points to determine more than
a single constant offset per ring with any precision.

The points in the lower panels in Figure \ref{figure5} show the errors
for the full noise model.  These are just the sum of the white noise
and `$1/f$' errors. The theoretical models, plotted as the solid lines,
are  simply equation (\ref{RT9}) rescaled to provide a good match to
the simulations, summed with the appropriate constant white
noise level. These provide an excellent match to the simulation results.
The general shape of the power spectrum errors $C^e_\ell$ plotted in Figures
\ref{figure5}c and \ref{figure5}f is well known from previous numerical work
on destriping (Delabrouille 1998; Maino \etals 1999; 
\etals 2000; Keih\"anen \etals 2003). However, the results presented here
explain why the errors have this particular form and how they depend on
the parameters of the experiment.
   
\begin{figure*}

\vskip 5.8 truein

\includegraphics{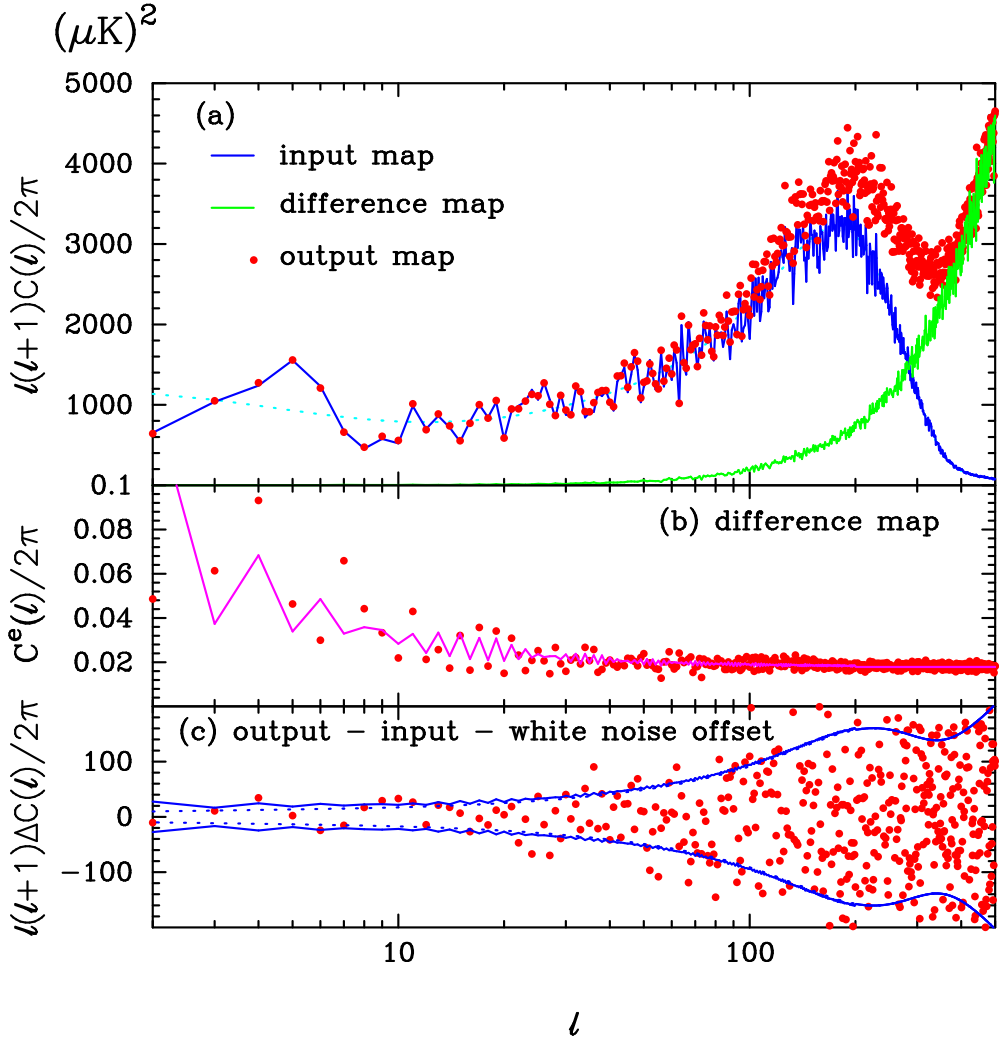}

\caption
{The top panel (a) shows the power spectrum of the input map (solid
line) and the power spectrum of the destriped map (filled
circles). The dashed line shows the power spectrum of the fiducial
$\Lambda$CDM model used to generate the input map. The power spectrum
of the difference (noise) map is also shown.  The middle panel (b)
shows the power spectrum of the difference map on an expanded scale
(filled circles) together with the error model shown in Figure
\ref{figure5}f.  The filled circles in the lower panel (c) show the
differences between the power spectrum of the destriped map. after
subtracting the constant white noise level, and the power spectrum of
the input map.  The solid lines show the dispersion expected from
equation (\ref{Sim1}) using the error model plotted in Figure (\ref{figure6}b).
The dashed lines show the disersion expected from white noise alone. }

\label{figure6}

\end{figure*}

Finally, in Figure \ref{figure6} we illustrate the effects of `$1/f$'
noise on a simulated map of the CMB sky for the scanning strategy with
a slow sinusoidal  precession. Figure \ref{figure6}a shows the power spectra of
the input map, the destriped map and for the difference map. The power
spectrum of the difference map is shown on a greatly expanded scale in
Figure \ref{figure6}b, together with the analytic model plotted in
Figure \ref{figure5}f. Figure \ref{figure6}c shows the differences
between the power spectra of the input and output maps but with the white noise
level subtracted, again plotted on an expanded scale. The solid lines show the expected
errors
\begin{equation}
     \langle (\Delta C_\ell)^2 \rangle^{1/2} = \sqrt{{(2(C^e_\ell)^2 + 4 C^e_\ell C_\ell) \over
(2 \ell + 1)}}, \label {Sim1}
\end{equation}
using the error model of Figure \ref{figure5}f. The dotted lines in this Figure show
the expected error from white noise alone. Notice that pure white noise is an excellent
approximation to the errors for $\ell \simgt 10$ and that even at low
multipoles the destriping errors are much smaller than the cosmic
variance $\langle \Delta C_\ell^2 \rangle = 2C_\ell^2/(2 \ell + 1)$. 
For example, the error in the quadrupole amplitude for this
realisation is $\Delta T_2^2 = \ell(\ell+1) \Delta C_2/2\pi = 10.4 \;{\rm \mu K^2}$, compared
to the expected cosmic variance of $\Delta T_2^2 = 717.8 \;{\rm \mu K^2}$. Compared to the
cosmic variance, the destriping errors at low multipoles can be ignored. Furthermore, 
errors of this magnitude are  much smaller than the errors of $50$-$100 \;{\rm \mu K}^2$
expected from inaccurate subtraction of the Galaxy ({\it cf} the discussion of 
the effects of Galactic subtraction on the low multipoles measured by WMAP, 
Bennet \etals 2003; Slosar and Seljak 2004).

\section{Conclusions}

In this paper, we have presented an analytic analysis of the effects
of destriping errors on the CMB power spectrum for various scanning
strategies. Destriping errors produce a characteristic error power
spectrum of the form shown in Figure \ref{figure3} which decays
roughly as $1/\ell$. The amplitude of the error power spectrum is
determined by the number of intersections in the ring set which
depends on the scanning strategy. In agreement with Stompor and White
(2003), and with earlier numerical work, there are so many
interconnections in a {\it Planck}-like scanning strategy that the
destriping errors should be small. Maps from {\it Planck} will
therefore be dominated by white noise, underneath which there will be
low amplitude correlated noise associated with striping. Nevertheless,
the low amplitude errors from striping introduce a characteristic
block-diagonal structure in the covariance matrix for the $a_{\ell m}$.
These  could  confuse searches for small amplitude physical effects,
such as non-Gaussian features of the CMB signal.

The implications of our analysis for the application of a hybrid power
spectrum estimator to {\it Planck}-like data are fairly
self-evident. Since the noise can be very accurately approximated as
white for multipoles $\ell \simgt 50$, the combination of a number of
pseudo-$C_\ell$ estimators with different pixel weighting schemes as
in E04 should give a close to optimal estimate of the power
spectrum at high multipoles. At low multipoles, an estimate of $C_\ell$ can be found by
applying a quadratic maximum likelihood (QML) estimator (Tegmark
1997c) to a low resolution map. This providesa close to optimal
estimate of the power spectrum at low multipoles, taking into account
of masked regions of the sky, and returns an estimate of the
covariance matrix $\langle C_\ell C_\ell^\prime \rangle$ that . For
the QML estimates, it should be an excellent approximation simply to
neglect destriping errors, since these are likely to be negligible
compared to the cosmic variance. If necessary, destriping errors can
be taken into account in the QML estimates, and folded into the
covariance matrix, by computing the full pixel-noise covariance matrix
on a low resolution map using an optimal map making algorithm.

Finally, this analysis has some implications for optimal map making algorithms.
The maximum likelihood map $\tilde {\bf m}$ is given by the well-known expression
\begin{equation}
\tilde {\bf m} = ({\bf P}^T {\bf N}^{-1} {\bf P})^{-1}{\bf P}^{T} {\bf
N}^{-1} {\bf t}, \label{C1}
\end{equation} 
where ${\bf N}$ is the noise covariance matrix $N_{ij} = \langle n_in_j \rangle$.
Since ${\bf N}$ is symmetric, it can be written as
\begin{equation}
{\bf N}  = {\bf W} {\bf \Lambda} {\bf W}^T, \label{C2}
\end{equation} 
where ${\bf W}$ is orthogonal and ${\bf \Lambda}$ is diagonal. The matrix
${\bf W}^T$ is a `prewhitening filter', since the noise matrix of the vector
${\bf n}^\prime = {\bf W}^T{\bf t}$ is diagonal. In terms of the prewhitened time-stream, equation
(\ref{C1}) can be written as,
\begin{equation}
 {\bf P}^T {\bf W} {\bf \Lambda}^{-1} {\bf W}^T {\bf P} \tilde {\bf m}
 = {\bf P}^{T} {\bf W} {\bf \Lambda}^{-1} {\bf W}^T {\bf t}. \label{C3}
\end{equation} 
Now consider equation (\ref{C3}) applied to a TOD consisting of
repeated scans of a single ring with stationary noise of an arbitrary
power spectral shape. If edge effects are ignored then the components
of ${\bf \Lambda}$ will be identical,  in which case the solution of
(\ref{C3}) is obviously the average of $t$ over the rings (equation
\ref{DS4}).  The {\it rhs} of equation (\ref{C3}) effectively filters
the TOD, but this is undone by the {\it lhs} to return the average of
the signal over the rings. 

Now consider a TOD consisting of a set of rings as in the {\it
Planck}-like scanning strategies considered in this paper.  The
maximum likelihood map will differ from a simple average over rings
because the rings cross. Thus optimal map making performs destriping
by comparing the crossing points of the TOD. However, since the noise
on a single ring is dominated by white noise, optimal map making
cannot reduce the errors much below the `irreducible' white noise
errors shown in Figures \ref{figure5}a and \ref{figure5}d. Optimal map
making can, in principle, reduce the map making errors associated with
`$1/f$' noise ({\it cf} Figures \ref{figure5}b and \ref{figure5}e) if
the knee frequency is significantly greater than the spin frequency,
but even in this case, the ability to reduce these errors will be
limited by the white noise on a ring. These arguments suggest that for
a {\it Planck}-like scanning strategy\footnote{The situation is
different for a WMAP-type strategy with fast precession (see Bennett
\etals 2003; Hinshaw \etals 2003), in which a pixel on the sky is
observed on many different timescales.}, simple destriping algorithms
will be very close to optimal, and may actually be preferable to
optimal algorithms because of their speed and because they require
fewer assumptions about the noise. Optimal algorithms are only optimal
if the noise model is accurate. In practice, with realistic
non-stationary noise, simple destriping may perform just as well and
conceivably better than an optimal algorithm.

\medskip

\noindent
{\bf Acknowledgements:} I thank members of CITA for their hospitality
during a visit where this work was begun. I am especially grateful to
Mark Ashdown, Dick Bond, Anthony Challinor and Floor van Leeuwen for
helpful discussions.


\begin{thebibliography}{}
\bibitem[\protect\citename{BKMR}2004]{BKMR}
Bartolo N., Komatsu E., Matarrese S., Riotto A.,
2004, submitted to Physics Reports. astro-ph/0406398.

\bibitem[\protect\citename{Betal03}2003]{Betal03}
Bennett, C. \etal, 2003, ApJS, 148, 1.

\bibitem[\protect\citename{BS93}1993]{BS93}
Brink D.M., Satchler G.R,. 1993, {\it Angular Momentum}, third edition, Oxford University
Press, Oxford.

\bibitem[\protect\citename{BFJS}2001]{BFJS01}
Borrill J., Ferreira P. G., Jaffe A. H., Stompor R., 2001, In {\it
`Mining the Sky'}, Proceedings of the MPA/ESO/MPE Workshop, Edited by 
A. J. Banday, S. Zaroubi, and M.Bartelmann. Springer-Verlag, Heidelberg,  p403.

\bibitem[\protect\citename{BMMDMBM97}1997]{BMMDMBM97}
Burigana C., Malaspina M., Mandolesi N., Danese L., Maino D., Bersanelli M.,
Maltoni M., 1997, Internal Report ITESRE. astro-ph/9906360.

\bibitem[\protect\citename{D98}1998]{D98}
Delabrouille, J.,  1998,  A\&A Suppl. Ser., 127, 555.

\bibitem[\protect\citename{DKP01}2001]{DKP01}
Dor\'e O., Teyssier R., Bouchet F.R., Vibert D., Prunet S., 2001,
A\&A, 374, 358.

\bibitem[\protect\citename{E03b}2003]{E04}
Efstathiou G., 2004,  MNRAS, 439, 603.

\bibitem[\protect\citename{BP86}1986]{GP86}
Groth E.J., Peebles P.J.E., 1986,  ApJ, 310, 507.

\bibitem[\protect\citename{Hetal03}2003]{Hetal03}
Hinshaw G. {\it et al.}, 2003,  ApJS, 148, 63.

\bibitem[\protect\citename{KKPMB03}2003]{KKPMB03}
Keih\"anen E., Kurki-Suonio H., Poutanen T., Maino, D., Burigana C., 
2003,  submitted to A\&A. astro-ph/0304411.

\bibitem[\protect\citename{Metal99}1999]{Metal99}
Maino D., {\it et al.}, 1999,  A\&A Suppl. Ser., 140, 383.

\bibitem[\protect\citename{MES96}1996]{MES96}
Maddox S.J., Efstathiou G., Sutherland W.,  1996, 
MNRAS, 283, 1227.

\bibitem[\protect\citename{NdeGGV01}1997]{NdeGGV01}
Natoli P., de Gasperis G., Gheller C., Vittorio N., 2001, A\&A, 372, 
346.

\bibitem[\protect\citename{PMKKH04}2004]{PMKKH04}
Poutanen T., Maino, D., Kurki-Suonio H., Keih\"anen E.,  Hivon E., 
2004,  submitted to MNRAS. astro-ph/0404134.

\bibitem[\protect\citename{RKACDK00}2000]{RKACD00}
Revenu B., Kim A., Ansari R., Couchot F., Delabrouille J., Kaplan J.,
2000,  A\&AS, 142, 499.

\bibitem[\protect\citename{SS04}2004]{SS04}
Slosar A.,  Seljak U., 2004, submitted to PRD, astro-ph/0404567.

\bibitem[\protect\citename{Setal03}2003]{Setal03}
Spergel D.N. \etal, 2003, ApJS, 148, 175. 

\bibitem[\protect\citename{SW4}2004]{SW04}
Stompor R., White M., 2004, A\&A, 419, 783.

\bibitem[\protect\citename{T97a}1997]{T97a}
Tegmark M., 1997a, ApJL, 480, L87.

\bibitem[\protect\citename{T97b}1997]{T97b}
Tegmark M., 1997b, PRD, 56, 4514.

\bibitem[\protect\citename{T97c}1997]{T97c}
Tegmark M., 1997c, PRD, 55, 5895.

\bibitem[\protect\citename{T03}1997]{T03}
Tuovinen J., 2003, Planck Newsletter Number 4., p7. 
(http://www.rssd.esa.int/SA/PLANCK/docs/Newsletters/PlanckNewsletter4.pdf)

\bibitem[\protect\citename{VMK88}1988]{VMK88}
Varshalovich, D.A., Moskalev A.N., Khersonskii V.K., 1988, {\it Quantum Theory of
Angular Momentum}, World Scientific, Singapore.

\bibitem[\protect\citename{vLetal}2002]{vLetal02}
van Leeuwen F., {\it et al.}, 2002, MNRAS, 331, 975.

\bibitem[\protect\citename{W96}1996]{W96}
Wright E.L., 1996, astro-ph/9612006.

\bibitem[\protect\citename{WHB96}1996]{WHB96}
Wright E.L., Hinshaw G., Bennett C.L.,  1996, ApJ, 458, L53.


\end{thebibliography}
\end{document}